\documentclass[preprint,nofootinbib]{revtex4}
\usepackage{graphicx}
\usepackage{graphicx,array,dcolumn}
\usepackage{calc,tabularx, epsfig,mathrsfs}
\usepackage{hyperref}
\usepackage{tikz}
\usepackage{amsmath,verbatim,enumerate}
\usepackage{amssymb}
\usepackage{multirow}

\usepackage{bm}
\usetikzlibrary{patterns}
\def \x{\mathbf{x}}
\def \a{\mathbf{a}}
\def \b{\mathbf{b}}
\def \O{\mathcal{O}}

\def\be{\begin{equation}}
\def\ee{\end{equation}}
\def \ell{\mathbf{\Lambda}}
\def\bal{\begin{align}}

\def\eal{\end{align}}
\begin{document}

\title{Subregion bulk reconstruction and symmetries }

\author{Nirmalya Kajuri}
\email{nkajuri@cmi.ac.in}
\affiliation{Chennai Mathematical Institute, Siruseri
Kelambakkam 603103}

\begin{abstract}
How do subregion boundary representations transform under conformal transformations? In this paper we conjecture a transformation rule and provide  evidence for it. We also discuss some consequences of this rule and show its connection to the subregion paradox given by Alhmeiri et al.

\end{abstract}

\maketitle

\section{Introduction.} 

The AdS/CFT conjecture \cite{Maldacena:1997re,Gubser:1998bc,Witten:1998qj} tells us that a quantum theory of gravity on a d+1 dimensional asymptotically Anti-de Sitter spacetime is equivalent to a d dimensional CFT living on the boundary of the asymptotically anti-de Sitter spacetime. This means that the bulk physics can in principle be 'reconstructed' from the boundary theory. 

The bulk reconstruction program is a step in this direction. It aims to find operators in the boundary CFT that represent the fields in the bulk\cite{Balasubramanian:1998sn,Banks:1998dd,Dobrev:1998md,Bena:1999jv,Hamilton:2005ju,Hamilton:2006az,Kabat:2011rz,Kabat:2012hp,Papadodimas:2012aq,Kabat:2012av,Kabat:2013wga,Morrison:2014jha,Sarkar:2014dma,Kabat:2017mun,Foit:2019nsr}. A recent review can be found in \cite{Kajuri:2020vxf}.

The HKLL prescription for finding boundary representations of bulk fields is to solve the wave equation in global or Poincare chart of AdS. They turn out to be non-local CFT operators of the form:

\be 
\phi(X) = \int \, dy\, K(X;y) \O(y)
\ee

Where $y$ is the boundary coordinate and $\O(y)$ is the CFT primary dual to the bulk field $\phi$. The function $K(X;y)$ is called the smearing function. It has support on all boundary points with spacelike separation from the bulk point $X$.

How do these non-local operators transform under conformal transformations? The answer is known (chiefly from \cite{Nakayama:2015mva}, see also \cite{Miyaji:2015fia,Nakayama:2016xvw,Goto:2017olq}). The transformation rule is such that consistency between boundary conformal symmetries and bulk isometries is maintained. It is given by:

\begin{equation} \label{equation}
U(\Lambda) \phi (X) U^{-1}(\Lambda) = \phi (\ell^{-1} X)
\end{equation}

where $\Lambda$ is any boundary conformal symmetry while $\mathbf{\Lambda}$ denotes the corresponding bulk isometry.

One can similarly find boundary representations of fields in an AdS/RIndler wedge by solving equations of motion in an AdS/Rindler wedge. This gives a nonlocal operator which is smeared on the boundary of that AdS/Rindler wedge. The smearing function $K(X,y)$ is a distribution in this case \cite{Morrison:2014jha}.

In is interesting to ask, how does an AdS/Rindler wedge representation of a bulk field transform under conformal symmetries? The AdS/Rindler wedge is not preserved under all isometries. Likewise the boundary of the AdS/Rindler wedge is not preserved under all conformal transformations. However an AdS isometry will map an AdS/Rindler wedge to anotherAdS/Rindler wedge, and on the boundary this isometry will act as a conformal transformation. 

Based on this one may expect that under a conformal transformation, the boundary representation of a bulk field for one wedge will get mapped to the boundary representation of a different wedge.

Let us consider an isometry $\mathbf{\Lambda}$ that maps one wedge to a different wedge. Let us denote the initial wedge as $\a$ and the corresponding boundary representation of a scalar field as $\phi_\a$. Under the isometry $\a$ is mapped to the wedge $\mathbf{\Lambda}^{-1} a$. The boundary representation for the same scalar field in this wedge is denoted as $\phi_{\mathbf{\Lambda}^{-1}\a} $.

We conjecture the following transformation law for AdS/Rindler boundary representations, for isometries (conformal symmetries) that don't preserve the wedge (boundary of the wedge):

\be \label{symmetry}
U(\Lambda) \phi_a(X) U^{-1}(\Lambda) = \phi_{\mathbf{\Lambda}^{-1}\a} (\mathbf{\Lambda}^{-1} X)
\ee

Here $\Lambda$ is the boundary conformal symmetry corresponding to the bulk isometry $\mathbf{\Lambda}$.

This is similar to the first transformation law, except that the two boundary  representations on LHS and RHS are different. They are the representations corresponding to the two AdS/Rindler wedges mapped by symmetry.

While a direct proof should exist, we have not been able to prove the law. However under a fairly weak assumption which is part of the code subspace conjecture, we can provide evidence for \eqref{symmetry} within the code subspace. 

We will then show some interesting consequences of \eqref{symmetry}.  We will see that it sheds some light on the subregion paradox presented in \cite{Almheiri:2014lwa} in that some of the simpler 'paradoxical' relations shown there directly follow from \eqref{symmetry}.

In the next section we present evidence for \eqref{symmetry}. In the third section we will discuss consequences that follow from \eqref{symmetry}. We conclude with a summary.

\section{Evidence for the transformation rule}

Before we start, let us recollect the paradox noted in \cite{Almheiri:2014lwa} and the resolution given therein.

The paradox is as follows: If we have two overlapping AdS/Rindler wedges $b$ and $b'$ and boundary representations of corresponding scalar fields are $\phi_b$ and  $\phi_{b'}$ we have the following equivalence:

\be \label{paradox}
\langle 0| \phi_b(x_1)....\phi_b(x_n)|0\rangle = \langle 0|\phi_{b'}(x_1)...\phi_{b'}(x_n)|0\rangle.
\ee

This above result gives rise to an apparent paradox. $\phi_b$ and $\phi_{b'}$ are two different CFT operators with different supports, but they both represent the same bulk field. There is thus an apparent redundancy of description of the bulk from the boundary --  a given bulk field can be described by more than one boundary operator. 
This appears paradoxical, because it seems one has multiple CFT operators  whose actions are identical. It would thus seem that the same information is somehow stored in the boundary CFT in different regions.

In \cite{Almheiri:2014lwa} it was suggested that this paradox may be resolved if the actions of the different representations only agree in a subspace of the full CFT Hilbert space. This is the subspace that encodes bulk effective field theory, and is known as code subspace. Thus Alhmeiri et al conjectured that

\be \label{code}
 \phi_b(x_1)....\phi_b(x_n)|\psi\rangle = |\phi_{b'}(x_1)...\phi_{b'}(x_n)|\psi\rangle.
\ee

They further conjectured that the way information is stored in holography parallels quantum error correction.

In this paper we will only assume \eqref{code} to prove our conjecture \eqref{symmetry}. Of course it may turn out that \eqref{symmetry} holds while \eqref{code} does not. We will hint at this possibility in the next section.

Let us consider an AdS/Rindler wedge $a$. The overlap of this wedge with the boundary is denoted as $A$. Let $\phi_a$ be the AdS/Rindler boundary representation of a bulk scalar in $a$ and $\phi$ is the global HKLL representation. Then $\phi_a$ has support on $A$ while $\phi$ has support on all points in the boundary spacelike separated from the point considered. We assume \eqref{code} to hold when one of the representations is the global HKLL representation (one can also obtain the proof using AdS/Rindler representations, but it becomes more complicated).

 According to the assumption made above, we will have for a bulk point $X$:  $$\phi_a(X)|\psi\rangle = \phi(X)|\psi\rangle$$. 

For any state $\psi$ in the code subspace. In what follows we will consider with the vacuum state, but the proof can be extended for states representing paricle excitations on the bulk: 

\begin{equation}\label{work}
\phi_a(X)|0\rangle = \phi(X)|0\rangle
\end{equation}

Now let $Y$ be another bulk point which is related to $X$ by a bulk isometry $\ell$: $Y = \ell^{-1} X$. We assume $\ell$ to be such that it \textit{does not} preserve the wedge, that is some points inside the wedge will be mapped out of it under this isometry. We call the corresponding boundary conformal transformation as $\Lambda$. This also does not preserve the boundary of the wedge.

Now from \eqref{equation} we know that $\phi(X)$ is mapped to $\phi(Y)$ via the unitary representation of the conformal transformation $\Lambda$: 

\begin{equation} 
U(\Lambda) \phi (X) U^{-1}(\Lambda) = \phi (Y).
\end{equation}

So we have: 

\be \label{dirt}
U(\Lambda) \phi (X)|0\rangle = \phi(Y)|0\rangle 
\ee

Applying the unitary operator $U(\Lambda)$ on both sides of \eqref{work} and using \eqref{dirt}, we have:
\be \label{righteous}
U(\Lambda) \phi_a (X)|0\rangle = \phi(Y)|0\rangle 
\ee

However we saw that the region $a$ is not preserved under this isometry. Let $a' = \ell^{-1}a$ be the AdS/Rindler wedge related to $a$ by this isometry. Let $A'$ is the boundary of $a'$, then we have $A' = \lambda^{-1}A$. 

What happens to $\phi_a$ under a conformal transformation?
We know that:
\be 
\phi_a (X) = \int_{A} dy\, K(X,y) \mathcal{O}(y)
\ee

The integration is over $A$. Then under a conformal transformation we have:
\be 
U(\Lambda)\phi_a (X)U^{-1}(\Lambda) = \int_{A} dy\, K(X,y)U(\Lambda) \mathcal{O}(y)U^{-1}(\Lambda) = \int_{A'}  \left|\frac{\partial y'}{\partial y'}\right|^{d-\Delta}\, d (\Lambda^{-1}y )\, K(X,y)  O(\Lambda^{-1}y)
\ee

So we have that $U(\Lambda)\phi_a (X)U^{-1}(\Lambda)$ is an operator smeared over $A'$. 

Now for the AdS/Rindler wedge $a'$ we will also have a boundary representation of the bulk scalar which we denote as $\phi_{a'}$ which is smeared over $A'$. As the bulk point $Y$ lies in this wedge, it follows again from the code subspace conjecture that:

\be \label{ri}
\phi_{a'}(Y)|0\rangle = \phi(Y)|0\rangle 
\ee

Then from \eqref{righteous} and \eqref{ri} it follows that:
\be 
U(\Lambda)\phi_a (X)U^{-1}(\Lambda)|0\rangle=\phi_{a'}(Y)|0\rangle
\ee

This we see that these two operators related by \eqref{symmetry} have the same action on the vacuum state. 

This can be extended to any state in the code subspace. Let's take any state $|\psi\rangle$ that belongs to the code subspace. It would represent some state in the bulk field theory (such as a many-particle excitation). The action of bulk isometry would map it to another state in bulk field theory. By consistency between bulk isometries and boundary conformal transformations, the boundary dual to this bulk state would be  $ \psi'\rangle =U(\Lambda)|\psi\rangle$. Thus this state also belongs to the code subspace.

Now from code subspace conjecture we  have:

\begin{equation} 
\phi_a(X)|\psi\rangle = \phi(X)|\psi\rangle
\end{equation}

Acting on $U(\Lambda)$ on both sides we have:
\begin{equation}\label{bwork}
U(\Lambda)\phi_a(X)|\psi\rangle =U(\Lambda) \phi(X)|\psi\rangle
\end{equation}

Using \eqref{equation} we have:
\be \label{bri}
U(\Lambda) \phi(X)|\psi\rangle = \phi(Y)|\psi'\rangle
\ee

Also since $|\psi'\rangle$ is a state in the code subspace, from the code subspace conjecture we have 
\be \label{bdirt}
 \phi_{a'}(Y)|\psi'\rangle = \phi(Y)|\psi'\rangle
\ee

Then using \eqref{bdirt} and \eqref{bwork} we have that:
\be  
U(\Lambda)\phi_a(X)U^{-1}(\Lambda)|\psi'\rangle =\phi_{a'}(Y)|\psi'\rangle 
\ee

Thus we have shown here is that the transformation law \eqref{symmetry} holds in the code subspace. 

Moreover we saw that the operators on the LHS and RHS have the same support. This strongly suggests that the two operators are one and the same and \eqref{symmetry} holds generally. 

Thus we have obtained evidence for the transformation rule \eqref{symmetry}. A direct proof not involving any assumptions should be possible as the subregion representations of bulk fields are known, and the question is if they transform into one another under boundary conformal transformations. This is complicated by the non-existence of the smearing function in this case, and we have not been able to find a direct proof. However we should emphasize that the validity of \eqref{symmetry} is independent of the assumption \eqref{code}.

$r^\Delta_A$

\be 
\phi_A(r_A,\x _A)= \int d^d \x _A K(r_A,\x _A ; \x '_A)   \O(\x'_A)
\ee

\section{Consequences of the transformation rule}

In this section we will explore some consequences of \eqref{symmetry}. In particular we will show that some of the simpler instances on the subregion paradox of \cite{Almheiri:2014lwa} that we described in the last section follow directly from \eqref{symmetry} and don't require further explanation.

Specifically, \eqref{symmetry} issufficient when the paradox is formulated in terms of correlators in two different wedges, as below:
\be 
\langle 0| \phi_b(x_1)....\phi_b(x_n)|0\rangle = \langle 0|\phi_{b'}(x_1)...\phi_{b'}(x_n)|0\rangle.
\ee

To see this, let the isometry mapping $\b$  to $\b'$ be $\mathbf{\Lambda}$, and the conformal transformation mapping their boundaries be $\lambda$. Then from the transformation law \eqref{symmetry} it follows that:

\be \label{one}
 \langle 0|\phi_{\b'}(x_1)...\phi_{\b'}(x_n)|0\rangle = \langle 0|\phi_{\b}(\mathbf{\Lambda}^{-1} x_1)...\phi_{\b}(\mathbf{\Lambda}^{-1} x_n)|0\rangle
\ee

From \eqref{paradox} to follow from \eqref{one}, we should have that:
\be \label{two}
\langle 0|\phi_{\b}(\mathbf{\Lambda}^{-1} x_1)...\phi_{\b}\mathbf{(\Lambda}^{-1} x_n)|0\rangle = \langle 0| \phi_\b(x_1)....\phi_\b(x_n)|0\rangle
\ee

This relation that says bulk isometry should to be respected by the boundary representation. It is incorporated in boundary representations by construction.  While we showed this for correlators in the vacuum, it is straightforward to extend it to other states in the code subspace. 

Thus we see that this simple case of the paradox can be directly understood as a consequence of the transformation rule \eqref{symmetry} and does not require further explanation.

\section{Summary and Conclusions}

In this paper we conjectured that the transformation of AdS/Rindler representations under boundary conformal symmetries is given by \eqref{symmetry}. Assuming that different representations have the same action on the code subspace we proved that this statement holds within that subspace and argued that it holds generally. 

Then we assumed \eqref{symmetry} to be true and explored some consequences. We showed that the examples of the subregion paradox involving only two wedges is a direct consequence of the transformation law-- same information can reside in different locations if they are related by symmetry. This is an interesting consequence of the transformation rule.

As we noted before, one should be able to prove the transformation law \eqref{symmetry} directly. We leave this for future work.

\acknowledgements

I would like to thank Alok Laddha, Ronak Soni and Aron Wall for helpful discussions.

\end{document}